\documentclass[draft]{mn2e}

\title[Infrared spectroscopy of Nova Cassiopeiae 1993]{Infrared spectroscopy of Nova Cassiopeiae
1993 (V705 Cas) -- IV. A closer look at the dust}
\author[A. Evans et al.]{A. Evans$^1${\thanks{Contact for correspondence. Email address {\tt
ae@astro.keele.ac.uk}}}, V. H. Tyne$^1$, O. Smith$^1$, T. R. Geballe$^2$, J. M. C. Rawlings$^3$,
\newauthor S. P. S. Eyres$^4$ \\
$^1$ Astrophysics Group, School of Chemistry \& Physics, Keele University, Keele, Staffordshire, ST5 5BG \\
$^2$ Gemini Observatory, 670 N. A'ohoku Place, Hilo, HI\,96720, USA \\
$^3$ Department of Physics and Astronomy, University College London, Gower Street, London, WC1E 6BT \\
$^4$ Centre for Astrophysics, University of Central Lancashire, Preston, PR1 2HE, UK}

\date{Version of 26 March 2005; ME1022rv.tex}

\pagerange{\pageref{firstpage}--\pageref{lastpage}}

\def\LaTeX{L\kern-.36em\raise.3ex\hbox{a}\kern-.15em
    T\kern-.1667em\lower.7ex\hbox{E}\kern-.125emX}

\newcommand{\pion}[2]{{#1}\,{\sc {#2}}}
\newcommand{\fion}[2]{[{#1}\,{\sc {#2}}]}
\newcommand{\speedclass}{\mbox{$\dot{m}_{\rm vis}$}}

\newcommand{\vunit}{\mbox{\,km\,s$^{-1}$}}

\newcommand{\Lsun}{\mbox{\,$L_\odot$}}
\newcommand{\mic}{\mbox{$\,\mu$m}}
\newcommand{\ncas}{\mbox{V705~Cas}}

\newcommand{\chemone}{\raisebox{0.03cm}{$-$}} 

\newcommand{\ltsimeq}{\raisebox{-0.6ex}{$\,\stackrel 
           {\raisebox{-.2ex}{$\textstyle <$}}{\sim}\,$}} 
\newcommand{\gtsimeq}{\raisebox{-0.6ex}{$\,\stackrel
           {\raisebox{-.2ex}{$\textstyle >$}}{\sim}\,$}}

\begin{document}

\label{firstpage}
\maketitle
\begin{abstract}
Nova Cassiopeiae 1993 (V705 Cas) was an archetypical dust-forming
nova. It displayed a deep minimum in the visual light curve, and
spectroscopic evidence for carbon, hydrocarbon and silicate dust.
We report the results of fitting the infrared spectral energy
distribution with the {\sc dusty} code, which we use to determine
the properties and geometry of the emitting dust. The emission is
well described as originating in a thin shell whose dust has a
carbon:silicate ratio of ~2:1 by number ($\sim1.26$:1 by mass) and a
relatively flat size distribution. The 9.7\mic\ and 18\mic\ silicate
features are consistent with freshly-condensed dust and, while
the lower limit to the grain size distribution is not well
constrained, the largest grains have dimensions $\sim0.06$\mic;
unless the grains in \ncas\ were anomalously small, the sizes of
grains produced in nova eruptions may previously have been
overestimated in novae with optically thick dust shells.
Laboratory work by Grishko \& Duley may provide clues to
the apparently unique nature of nova UIR features.
\end{abstract}

\section {Introduction}

\ncas\ (1993) was a typical dust-forming nova which showed evidence
for carbon, silicate and hydrocarbon dust (Evans et al. 1997, Mason 
et al. 1998). Early IUE observations (Shore et al. 1994) suggested
that the grains grew to $\sim0.2$\mic\ shortly after dust condensation.
Mason et al. (1998) modelled broadband infrared (IR) data using a
combination of dust types, and concluded that the circumstellar
dust shell consisted of carbon and silicate grains in the ratio
$\sim8.5$:1 by mass.

Evans et al. (1997; hereafter Paper~II) presented IR spectroscopy
of the dust in \ncas, with some data covering the wavelength
range 2--24\mic. In addition to nebular and coronal features,
emission features normally associated with hydrocarbons
-- the so-called `Unidentified Infrared' (UIR) bands -- were
present in the spectra. The UIR features in \ncas\  were at
wavelengths 3.28, 3.4, 8.1, 8.7 and 11.4\mic; the `8.1' and
`11.4' feature seem to correspond respectively to the UIR
features normally seen at 7.7\mic\ and 11.25\mic\ (but see
\S\ref{UIRf} below), while the `3.4' feature was much stronger
relative to the `3.28' than is generally the case in other
astrophysical environments. The properties of the UIR bands
in \ncas\ and in other novae indicate a set of UIR features
peculiar to novae (Geballe 1997), due either to the nova-specific
environment or excitation conditions.

The 10\mic\ spectrum of \ncas\ also displayed a prominent silicate
feature at $\sim10$\mic\ and possibly a broad, weak feature at
18\mic\ (Paper~II). The 10\mic\ feature in \ncas\ was narrow,
and peaked at 9.7\mic, in contrast to the broader silicate
features in novae Aql 1982 and Her 1991, which were also shifted
to longer wavelength (see Smith et al. 1995). This implies that
the silicate in novae Aql and Her displayed a degree of crystallinity,
whereas that in \ncas\ was amorphous.

\begin{figure*}
\begin{center}
\setlength{\unitlength}{1cm}
\begin{center}
\leavevmode
\begin{picture}(6.0,5.5)
\put(0.0,4.0){\includegraphics{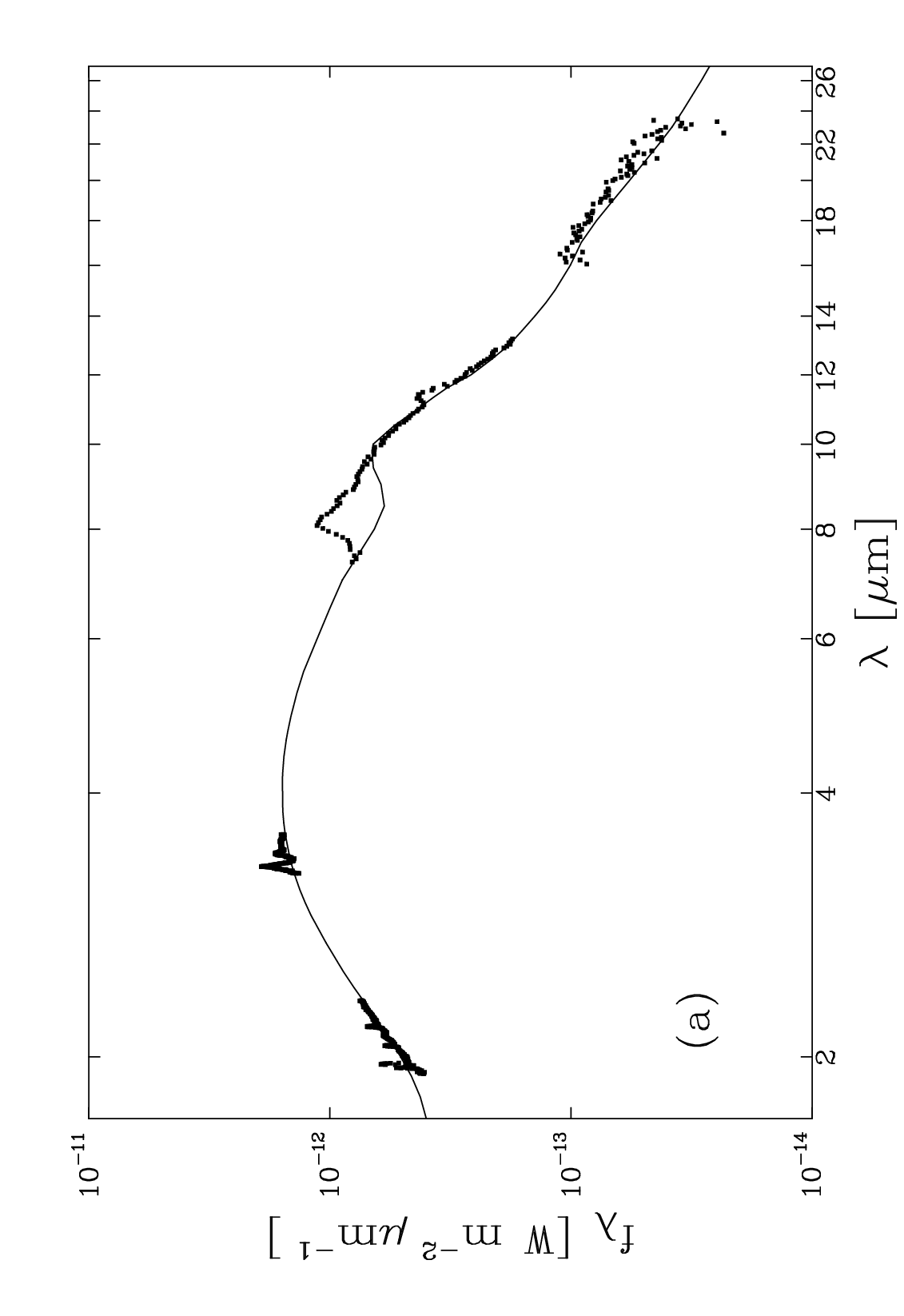}}
\put(0.0,4.0){\includegraphics{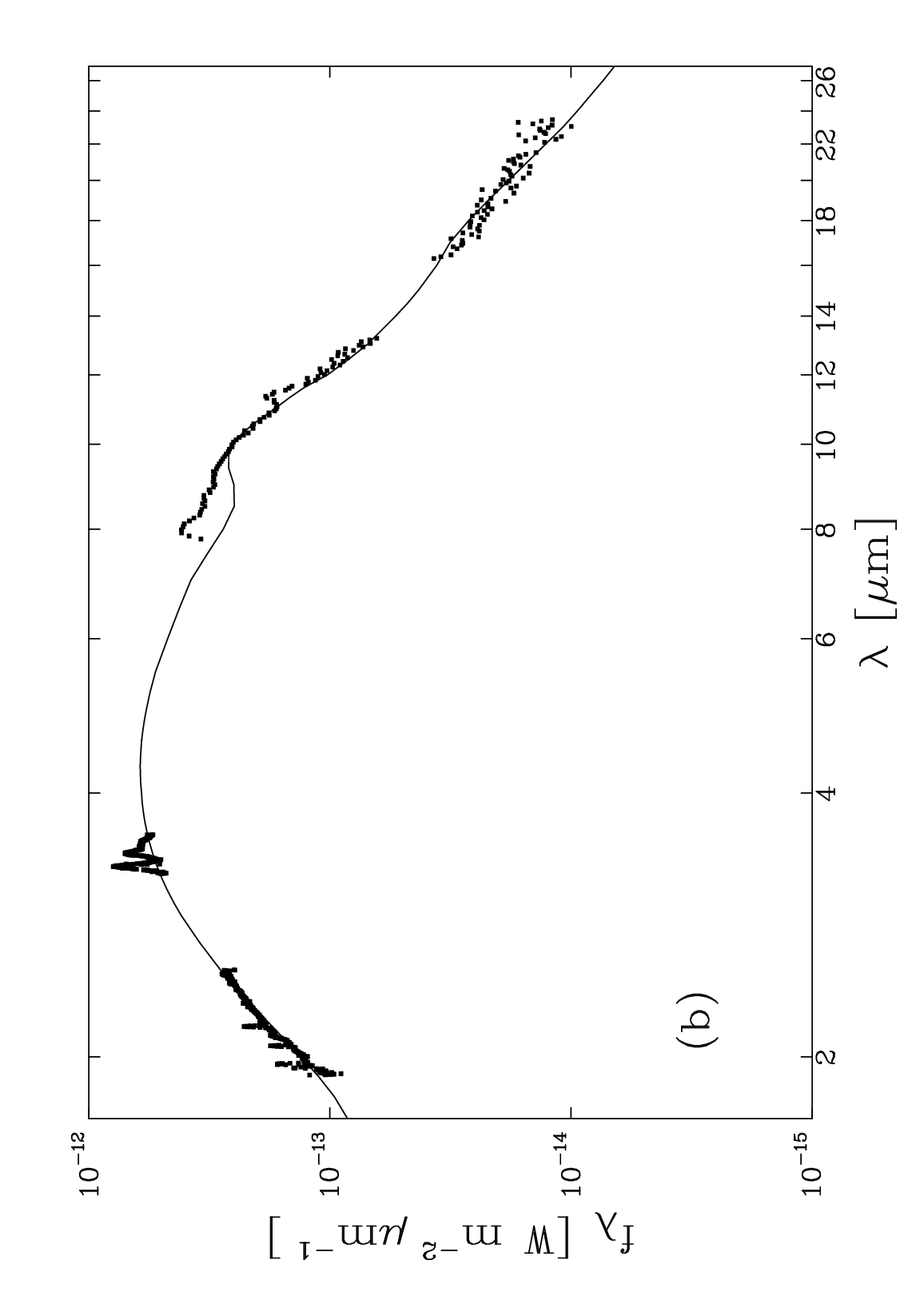}}
\end{picture}
\end{center}
\caption[]{(a) Fit of {\sc dusty} model to 1994 August data for nova
           \ncas; thick curve is {\sc dusty} model. Note the strong UIR
           features in the 10\mic\ window; the apparent `noise' in the
           $\sim2$\mic\ window is caused by the nebular lines. (b) As
	   (a) but for 1994 October/November data.}
\label{sed}
\end{center}
\end{figure*}

In Paper~II we fitted a simple function of the form
$\nu^{\beta}\,B(\nu,T)$, where $B$ is the Planck function
at temperature $T$ and $\beta$ ($\simeq1$) is a constant (the
so-called $\beta$-index for the dust), to IR spectra in the range
2--24\mic, and concluded that the grains eventually grew to
$\sim0.7$\mic. However, given the high quality of the data and the
availability of the {\sc dusty} code -- which solves the radiative
transfer in a dusty environment for a variety of conditions
(Ivezi\'{c} \& Elitzur 1997, Ivezi\'{c}, Nenkova \& Elitzur 1999)
-- we are now in a position to make a more sophisticated attempt
at fitting the data and to re-evaluate the nature of the dust.

Here we describe the application of {\sc dusty} to model the
dust around \ncas.

\section{The data}

The data analysed here are described in detail in Paper~II; for
consistency with earlier work we take reddening $E(B-V)=0.5$
(Hauschildt et al. 1994) and 1993 December 14 as the origin of
time. We revisit the distance below.

The fitting procedure works best if there are data on either
side of the peak of the dust emission. For \ncas, two datasets
satisfy this criterion, namely those for 1994 August (to which
we shall refer as Epoch~1), and 1994 October/November (Epoch~2);
at these two epochs we have quasi-simultaneous data over the
wavelength range 2--4\mic\ ($K,L$), 7.5--13\mic\ ($N$), 16--24\mic\
($Q$), at resolution $\sim300-1000$ ($K,L$ bands) and $\sim60$
($N,Q$ bands). The Epoch~1 data were obtained within 4~days of
each other (on days 251 and 255 of the outburst, from Paper~II;
we take $t=253$~days), while the Epoch~2 data were obtained within
40~days of each other (days 300 and 341; we take $t=320$~days).
The data are shown in Fig.~\ref{sed}.

In each case we assume that the nova and its environment did not
change substantially between the times the data were obtained.
Obviously this assumption is less secure for the October/November
data but our attempts in Paper~II to fit the data with the simple
function as discussed above suggest that this assumption is not
unreasonable.

\begin{figure*}
\begin{center}
\setlength{\unitlength}{1cm}
\begin{center}
\leavevmode
\begin{picture}(6.0,5.5)
\put(0.0,4.0){\includegraphics{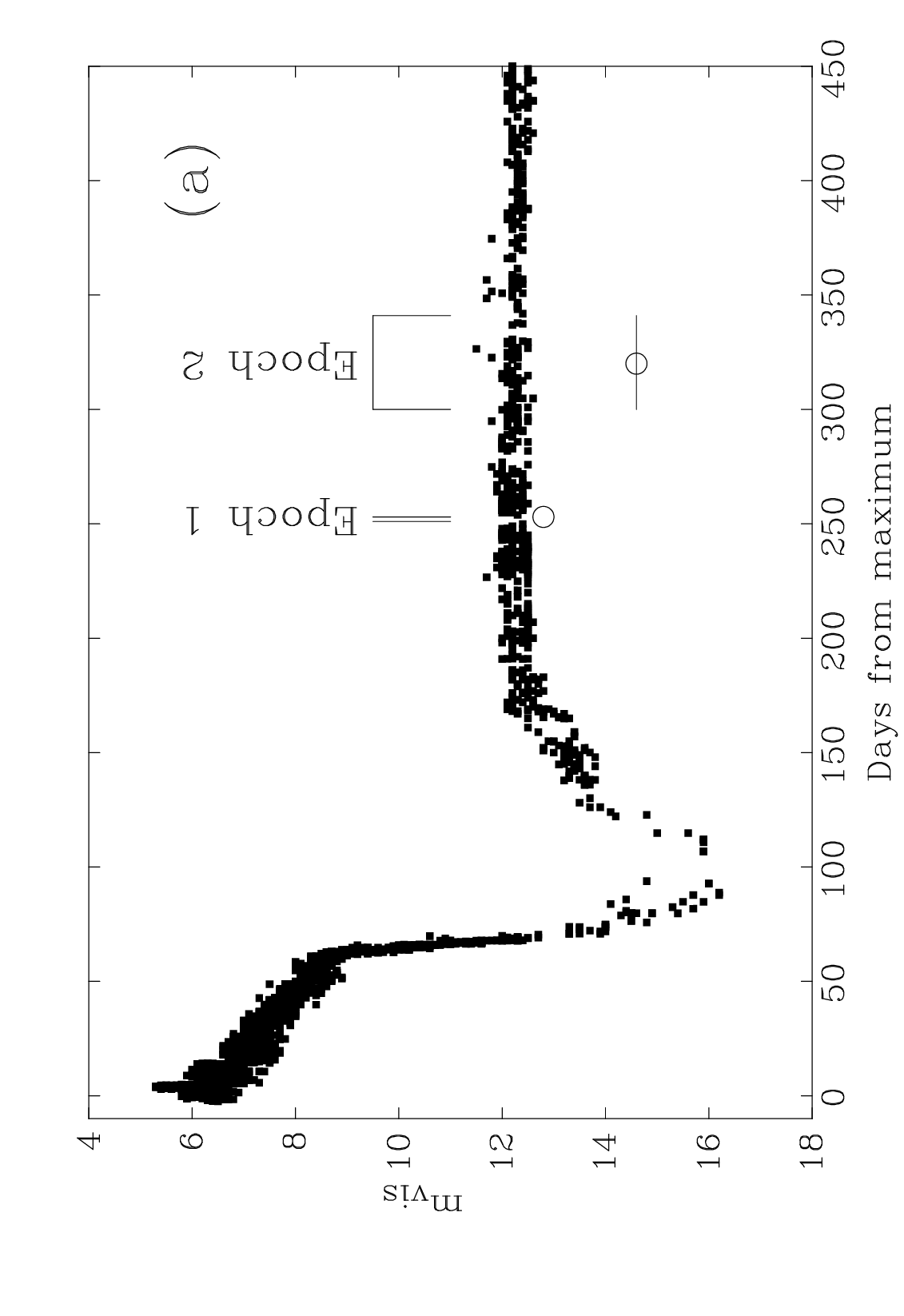}}
\put(0.0,4.0){\includegraphics{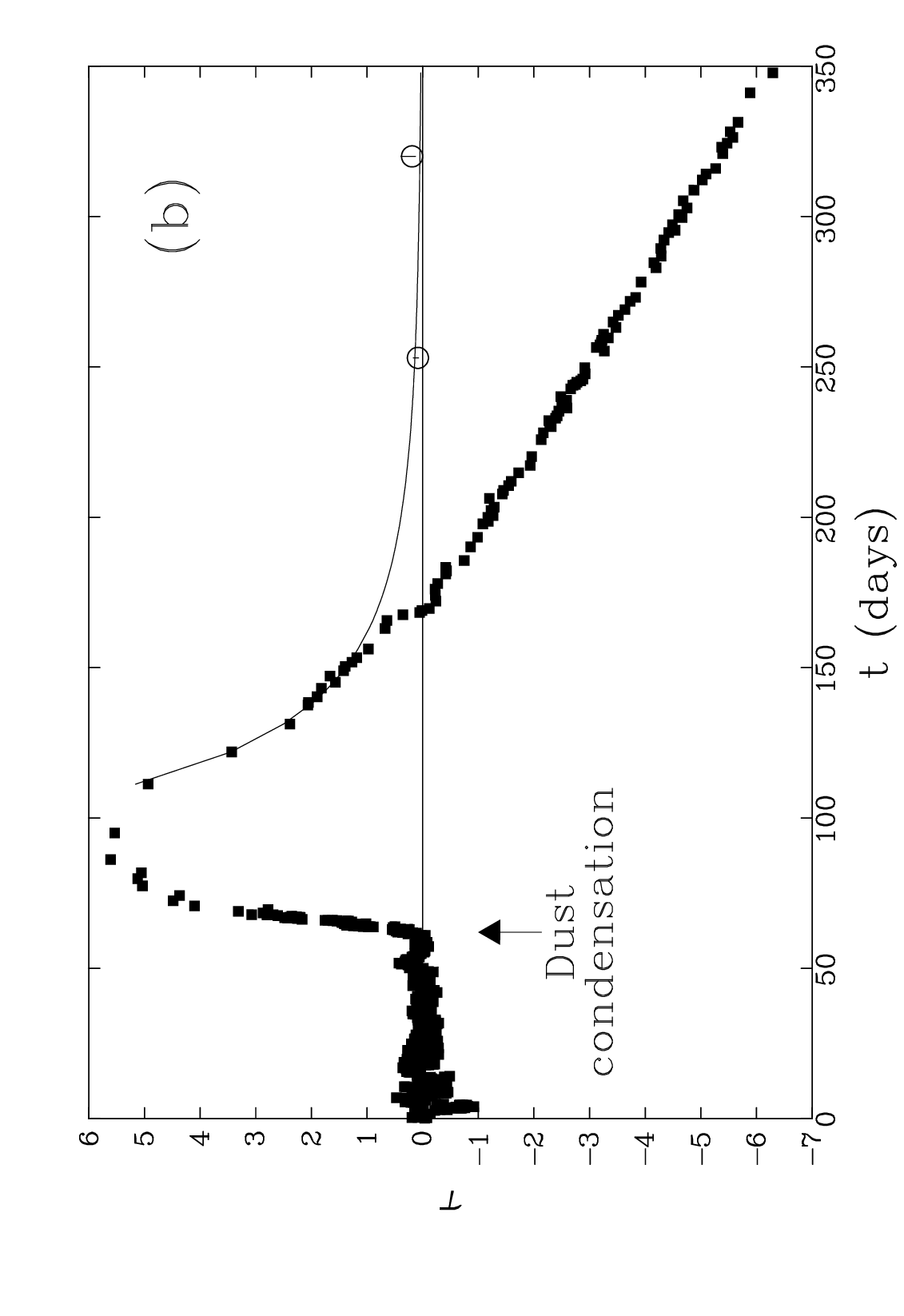}}
\end{picture}
\end{center}
\caption[]{(a) Visual light curve of \ncas\ (filled squares), with
Epochs 1 and 2 indicated; open circles are estimated location of
pseudo\-photospheric continuum, based on {\sc dusty} fits and after
applying interstellar extinction corresponding to $E(B-V)=0.5$.
(b)~Time-dependence of the optical depth in the dust shell during
light curve recovery, estimated as described in text. This is
effectively the visual light curve of \ncas\ but rectified by
Equation~(\ref{lightcurve}), which is represented by the horizontal
line. The curve is an estimate of the dust optical depth after the
deep minimum; open circles are values of $\tau$ from {\sc dusty} fits.
See text for details.}
\label{tau1}
\end{center}
\end{figure*}

\section{The fitting}

\subsection{The state of \ncas\ in 1994}

\subsubsection{Preamble}
\label{preamble}
We use observations of \ncas\ to estimate the likely
properties of the nova and its environment at the time of our
observations. For the purposes of the {\sc dusty} fitting we
shall simply assume that the stellar remnant radiates like
a blackbody. While more sophisticated models of the nova
photosphere are now routinely available (see Hauschildt et
al. 2002 and references therein), we have no {\it a priori}
information about the state of ionization of the ejecta;
however we expect that they are transparent to photons in the
Lyman continuum at the times of interest.

The effective temperature $T_*$ of a nova radiating at constant
bolometric luminosity rises as the pseudo\-photosphere
collapses back onto the white dwarf (Bath \& Shaviv 1976).
Consequently the emission shifts to shorter wavelengths and
the visual flux (in mag) declines according to
\begin{equation}
|\Delta m_{\rm vis}| \simeq 7.5 \: \log_{10} \left ( \frac{T_*}{T_0}
            \right )   \:\: ,
\label{deltam}
\end{equation}
where $\Delta m_{\rm vis}$ is the decline in mag from maximum
light and $T_0\simeq15,280$~K (Bath \& Shaviv 1976). However,
observational evidence is that novae at maximum are on the whole
rather cooler than 15,280~K, a value closer to $\simeq8,000$~K
being more typical (Beck et al., 1995; Warner 1995); indeed the
lower value is very close to the effective temperature determined
for \ncas\ around maximum by Hauschildt et al. (1994). We take
$T_0=8,000$~K here and use Equation~(\ref{deltam}) to
estimate $T_*$.

We have taken the light curve in Paper~II (reproduced in
Fig.~\ref{tau1}a) and determined \speedclass\ by a least
squares fit over the period $t=0$~days to $t=60$~days,
omitting `glitches' such as the small pre-dust `dip' noted
in Paper~II. We find
\begin{equation}
m_{\rm vis} = 6.32 (\pm 0.04) + 0.0364 (\pm0.0012) \,\, t \:\: ,
\label{lightcurve}
\end{equation}
where $t$ is in days; we therefore have that 
$\speedclass = 0.0364\pm0.0012$~mag~day$^{-1}$ ($t_3=82.4\pm2.7$~days,
$t_2=55.0\pm3.6$~days, $t_n$ being the time for the visual light curve
to decline by $n$~mag), which differs somewhat from that assumed
in Paper~II ($\speedclass=0.044$~mag~day$^{-1}$, $t_3\simeq 68$~days)
and is more in line with that given by Mason et al. (1998). We take
$t_3=82.4$~days here.

Using the usual relationship between absolute magnitude and rate of
decline (Della Valle \& Livio 1995), the absolute visual magnitude
of \ncas\ is $M_{\rm V}=-7.05$ at maximum light; using a bolometric
correction corresponding to an 8,000~K supergiant the corresponding
bolometric luminosity is $5.6\times10^4$\Lsun. The value of
$M_{\rm V}$, together with the value of $m_{\rm vis}(t=0)$ from
Equation~(\ref{lightcurve}) and the assumed reddening, gives a
distance of $D=2.3\pm0.4$~kpc (cf. 3~kpc assumed in
Paper~II), in which the uncertainty in the reddening is the major
contributor to the error.

\subsubsection{Effective temperature of the remnant}
\label{remtemp}
We estimate the effective temperature of the stellar remnant
using  Equation~(\ref{deltam}). For Epoch~1, we find
$T_* \simeq 135,000$~K, while for Epoch~2,
$T_* \simeq 285,000$~K; there is a $\sim10$\% uncertainty in $T_*$
that follows from the uncertainty in the coefficient of $t$ in
Equation~(\ref{lightcurve}). Although there are in principle
several ways of corroborating these values, complementary data
at other wavelengths are generally lacking.

In Paper~II we noted the presence of \fion{Si}{vi}
($^2$P$_{1/2}-^2$P$_{3/2}$) at $\lambda=1.963$\mic, which was fairly
strong at Epoch~1 and also present at Epoch~2 (\fion{Si}{vii} at
$\lambda=2.48$\mic\ was also present, but very weak, at Epoch~2).
The ionization potential of Si$^{4+}$ is $\sim167$~eV. If we
assume that the gas is photo\-ionized rather than collisionally
ionized (see Benjamin \& Dinerstein (1990) and Evans et al. (2003)
for a discussion of this), the presence of the \fion{Si}{vi}
feature implies a sufficient number of photons with energy
$\ge167$~eV. For strong \fion{Si}{vi} emission we suppose for the
purpose of estimate that half the emitted photons should be capable
of ionizing Si$^{4+}$, implying an effective temperature
$T_*\simeq1.93\times10^6$~K. At $T_*\simeq 285,000$~K, only
$\sim2$\% of the emitted photons are capable of ionizing
Si$^{4+}$. The presence of the
\fion{Si}{vi} coronal line therefore seems to imply a $T_*$
that is very much greater than 285,000~K; alternatively, a
different ionization mechanism may be operating.

\begin{table*}
\begin{center}
	\caption{Nova and dust shell parameters for \ncas\ for the two epochs. See
text for discussion.}
\label{dustyfit}
\begin{tabular}{lcccccc}
\hline
Property   & Epoch 1 & Epoch 2  \\ \hline
$t$ (days) & 253     & 320      \\
 &   &     \\
$F(\mbox{AC})$     & $0.64^{+0.10}_{-0.04}$   &  $0.70^{+0.19}_{-0.07}$     \\
 &   &     \\
$F(\mbox{Silicate})$  & $0.36^{+0.10}_{-0.04}$   &  $0.30^{+0.19}_{-0.07}$   \\
 &   &     \\
$T_{\rm d}$ (K)  & $604^{~+3}_{-16}$     &  $612^{+12}_{-24}$    \\
 &   &     \\
$a_{\rm min}$ ($\!\mic$)  & $0.003\pm0.013^*$ & $0.005\pm0.025^*$  \\
 &   &     \\
$a_{\rm max}$  ($\!\mic$) & $0.061\pm0.008^*$ & $0.055\pm0.018^*$ \\
 &   &     \\
$q$              & $2.3\pm0.5^*$      & $2.3\pm0.5^*$    \\
 &   &     \\
 $R_{\rm in}$ (cm)  &  $7.18^{+0.51}_{-0.58}\times10^{15}$ & $7.36^{+0.76}_{-1.37}\times10^{15}$ \\
 &   &     \\
 $\theta_{\rm in}$ (mas) & $417.3^{+29.6}_{-33.7}$ & $427.8^{+44.2}_{-79.6}$ \\
  &   &     \\ 
$1+\Delta R/R$     & $1.11\pm0.04^*$    & $1.24\pm0.19^*$    \\
 &   &     \\
$\tau$           & $0.085^{+0.082}_{-0.012}$     & $0.19^{+0.20}_{-0.06}$    \\  
 &   &     \\
$f_{2.3}$ ($10^{-10}$ W\,m$^{-2}$)       & 3.26  &  3.26    \\
		    &   &  \\
$f_{\rm dust}$ ($10^{-12}$ W\,m$^{-2}$)
                 & 13.1  &  4.34  \\ 
		    &   &  \\
$f_{\rm dust}/f_{2.3}$ & 0.040 & 0.013  \\
 &   &     \\
$T_*$ (K)        & 135,000  & 285,000  \\
 &   &    \\
Estimated $V_{\rm phot}$   & 12.8     & 14.8    \\ \hline
\end{tabular}
\end{center}
\end{table*}

On the other hand, Orio, Covington \& \"{O}gelman (2001) have
looked at {\sl ROSAT} \/ data for over 100 classical and recurrent
novae. Although \ncas\ is not included in their discussion, they
conclude that the `super-soft x-ray' phase of the nova outburst
is relatively short-lived, and is observed for only $\sim20$\%
of novae: hot post-nova white dwarfs are relatively rare. Orio
et al. hence estimate a general $1\sigma$ upper limit
$T_*<300,000$~K; the x-ray evidence (such as it is) seems to
point to a value of $T_*$ consistent with that implied by
Equation~(\ref{deltam}), and somewhat lower that that implied
by the coronal lines.

In our preliminary investigations (Evans et al. 2002) we
approximated the star by a blackbody with $T_*=70,000$~K, and
noted that the model IR spectrum is not sensitive to the
temperature of the central object. However the predicted flux
at $\lambda=0.55$\mic\ must not, after allowing for interstellar
extinction, exceed the value implied by the visual light curve.
This requires that the $m_{\rm vis}$ as calculated by {\sc dusty}
must be $\gtsimeq11$, which provides a further constraint on $T_*$.

\subsubsection{Dust shell thickness and optical depth}
\label{thickness}
CO was present in the IR spectrum of \ncas\ before maximum
(Evans et al. 1996, hereafter Paper~I). In Paper~I we
assumed that the relative thickness of the CO-bearing shell,
$\Delta R/R$ (where $R$ is the inner shell radius), was
$\sim0.1$ in \ncas. Furthermore consideration of the
dust-forming potential of novae, together with the rapid
recovery of the visual light curve after deep minimum, supports
the view that the dust too is confined to a thin shell (Rawlings
\& Evans 2005). The value of $\Delta R/R$ for the dust shell
is provided by our fitting procedure and, as the chemistry
that leads to grain nucleation and grain formation is expected
to be confined to the CO-bearing region, the {\sc dusty} fitting
may provide {\it a posteriori} justification for the assumption
$\Delta R/R\sim0.1$.

Dust precipitation began on day~62, after which the visual
light curve went into deep decline. At this time, the difference
between the {\em observed} visual magnitude and that predicted by
Equation~(\ref{deltam}) is a measure of the visual optical depth
$\tau$ of the dust shell; the time-dependence of $\tau$ is
plotted in Fig.~\ref{tau1}b. This is the visual light curve
of Paper~II, rectified by Equation~(\ref{lightcurve}), so
that the visual decline described by Equation~(\ref{lightcurve})
is represented by the horizontal line. As the dust shell
disperses (or breaks up) and the light curve recovers, the
pseudo\-photospheric emission shifts to the ultraviolet (UV);
at this time most of the visual light is provided by strong
emission lines and Equation~(\ref{deltam}) no longer applies;
for this reason the behaviour of the optical depth `data' in
Fig.~\ref{tau1}b for $t\gtsimeq165$~days is not meaningful.

We see from Fig.~\ref{tau1} that maximum dust optical depth in the
visual, $\tau_{\rm max} \simeq 5.5$, was attained at $t\simeq 90$~days,
and we estimate that the optical depth for Epoch~1 was $\sim0.1$,
for Epoch~2 $\sim0.05$. The precise values are not important at
this stage; however our fitting procedure determines the
optical depth in the dust shell and our analysis of the data
using {\sc dusty} should produce values of this order.

\subsubsection{Dust composition}

The presence of the UIR hydrocarbon features is a strong indication
that a form of carbonaceous dust is present. We therefore assume
that the carbon component is amorphous carbon (AC), for which the
optical constants are taken from Hanner (1988). As indicated above,
there is also a silicate component, and we take optical constants
for warm silicate from Ossenkopf, Henning \& Mathis (1992).

We use our fitting procedure to determine the relative amounts
of amorphous carbon and silicate dust in the shell.

\subsection{Fitting the {\sc dusty} models}
\label{fitting}
For a given central source effective temperature, dust composition
and dust shell properties, {\sc dusty} calculates the radiative
transfer through a spherically symmetric dust shell and determines
the observed spectral energy distribution (SED).
{\sc dusty} exploits the `scale-independence' of the radiative
transfer problem, so that the observed SED is effectively determined
only by optical depth in the dust shell; this means that `absolute'
values (e.g. luminosity and dust shell dimensions) are not uniquely
determined by the transfer problem and must be inferred by other
means.
We use a downhill simplex
routine (see Press et al. 1992) to explore the multi-dimensional
parameter space and to determine the best fit between the data and
the {\sc dusty} output (see Tyne et al. 2002 for details).

We assume an AC/warm silicate mix with a $r^{-2}$ dust density
distribution, appropriate for a steady wind. We optimize the fit
by varying the carbon:silicate ratio, the dust temperature at
the inner edge of the dust shell, the grain size distribution,
and the geometric and (visual) optical thickness of the dust
shell. Spectral features, namely the UIR features, and nebular
and coronal lines, were removed before fitting the {\sc dusty}
output to the data, and restored once the fit had been optimized.
The values of $t$, $f_{2.3}$ (the bolometric flux of the stellar
remnant normalized to 2.3~kpc) and $T_*$ for each epoch were fixed.

Further, as the relative placement of the spectra in the $K$, $L$, $N$
and $Q$ bands is uncertain by $\sim10-20$\% (a consequence of the
uncertainty in the flux calibration), the relative placements of the
individual spectral components were adjusted in an iterative way to
optimize the fit. For each $K\!LNQ$ dataset, {\sc dusty} was called
typically about 50--60 times per optimization; including the iteration
of the individual $K\!LNQ$ bands, {\sc dusty} ran some 300--400
times for each epoch to optimize the fit.

\section{Results \& discussion}
\label{results}

The best fits to the data for the two Epochs are shown in
Fig.~\ref{sed}a,b; these are equivalent to Figure~3a,b of
Paper~II except that in the latter, there was no attempt to
optimize the band-to-band fit (see \S\ref{fitting}), and the
peak of the `8.1' band was omitted.
The best fit parameters are listed in Table~\ref{dustyfit}, in
which $F(\mbox{AC})$ and $F(\mbox{Silicate})$ are respectively
the fractions (by number) of amorphous carbon and silicate grains
in the dust shell; by definition, $F(\mbox{AC}) + F(\mbox{Silicate})
\equiv 1$. $T_{\rm d}$ is the dust temperature at the inner edge of
the dust shell, $a_{\rm min}$ and $a_{\rm max}$ are the minimum
and maximum grain size in the grain size distribution
\[ n(a) \, da \propto a^{-q} \, da \:\:\: 
          (a_{\rm min} < a < a_{\rm max}) \: ; \]
here $a$ is grain radius, $n(a)\,da$ is the number of grains with
radius $\in[a, a+da]$, and $q$ is a constant.

In Table~\ref{dustyfit}, $f_{\rm dust}$ is the power output from
the dust shell (at 2.3~kpc), and $\tau$ is optical depth in the
$V$ band. For distance 2.3~kpc, the bolometric luminosity of the
nova is $5.6\times10^4$\Lsun\ at both Epochs (see Table~\ref{dustyfit}
and \S\ref{preamble}), a consequence of our assumption of constant
bolometric luminosity (see Equation~(\ref{deltam})). 

The uncertainties in a given parameter were obtained by fixing all
other parameters and varying the given parameter so as to increase
the reduced $\chi^2$ by unity on either side of the minimum;
where this procedure failed to
converge in a reasonable number of iterations and only one error
bound was obtained, this error is used and is indicated by an
asterisk in Table~\ref{dustyfit}. The values of $R_{\rm in}$
follow from the input radiation field and fitted $T_{\rm d}$ (see
Ivezi\'{c} et al. 1999); the uncertainties in $R_{\rm in}$ are derived
from the corresponding errors in $\theta_{\rm in}$.

Also given, in Table~\ref{dustyfit} and on Fig.~\ref{tau1}, is the
estimated contribution $V_{\rm phot}$ of the pseudo\-photosphere
to the observed visual magnitude, based on the pseudo\-photospheric
temperature, $\tau$ and interstellar extinction; the estimated
values of $V_{\rm phot}$ clearly satisfy the requirement that
they lie below the visual light curve (see \S\ref{remtemp}).

\subsection{The dust temperature}
\label{temp}
The dust temperature at the inner radius of the dust shell is
determined to be $T_{\rm d} = 604$~K at Epoch~1, and 612~K at
Epoch~2, i.e. essentially constant between the two Epochs within
the uncertainties. These values are comparable to those determined
by Mason et al. (1998, their Figure~1), on the basis of fitting
Planck functions to broadband data for $t\ltsimeq275$~days. However
if the central star delivers constant power input to the inner
radius of the dust shell, which consists of a single grain type
that is unchanging (so that the grain properties are also unchanging),
we must have that
\begin{equation}
 T_{\rm d} = T_{\rm c} \:\: \left ( \frac{t_{\rm c}}{t}
         \right )^{2/(\beta+4)} \:\: , 
\label{tcond}
\end{equation}
for a dust shell moving away from the site of the explosion at
uniform speed; here $T_{\rm c}$ is the grain condensation
temperature and $t_{\rm c}$ ($\simeq62$~days; see Fig.~\ref{tau1}a,b)
is the condensation time. From Equation~(\ref{tcond}), we expect
the value of $T_{\rm d} \, t^{2/(\beta+4)}$ to be $\simeq
\mbox{constant}$, whereas for \ncas\ it rises from
$\sim5524^{+28}_{-146}$~K~day$^{0.4}$ (Epoch~1) to
$\sim6149^{+121}_{-241}$~K~day$^{0.4}$ (Epoch~2), assuming
$\beta=1$ (the value assumed is not critical). This behaviour,
in which the dust temperature does not decline in accord with
Equation~(\ref{tcond}), is typical of dusty novae displaying
`isothermal' behaviour.

\subsection{The optical and geometrical depth}
\label{depth}
The geometrical thickness of the dust shell, $\Delta R/R$, is
$\simeq 0.11\pm0.04$ at Epoch~1, and $\simeq0.24\pm0.19$ at Epoch~2
(see Table~\ref{dustyfit}). In Paper~I we assumed that the relative
thickness of the CO-bearing shell, $\Delta R/R$ (where $R$ is the
inner shell radius), was $\sim0.1$ in \ncas. While this choice was
somewhat arbitrary it was governed by the fact that, in a nova wind
that declines with time, there will be a thin dense shell at the
outer edge of the ejecta (cf. Kwok 1983). Our {\sc dusty} fitting
indicates that the dust did indeed seem to have been confined to
a thin ($\Delta R/R\sim 0.1$) shell at the outer edge of the ejecta.
Indeed, the similarity of the dust temperatures at the inner shell
determined here (see \S\ref{temp}), and those determined by Mason
et al. (1998) on the assumption of an isothermal dust shell, is
a consequence of the fact that the dust shell {\em is} geometrically
thin so that the isothermal approximation is a good one.

{\em Very} roughly, for an optically thin dust shell in which grains
scatter isotropically, the ratio of the dust flux to the bolometric
stellar flux is $\sim$ the optical depth through the shell, suitably
averaged over wavelength. However the carbon and silicate grains
in \ncas\ do not scatter isotropicically and so this relationship will
only be approximately correct; nevertheless it is gratifying to see
in Table~\ref{dustyfit} that $f_{\rm dust}/f_{2.3} \sim \tau$.

\subsection{Dimensions of the dust shell}

The inner radius of the dust shell, $R_{\rm in}$, is
$7.18\times10^{15}$~cm ($7.36\times10^{15}$~cm) at Epoch~1
(Epoch~2). At 2.3~kpc, the corresponding angular diameter
$\theta$ of the inner edge of the dust shell is 417~mas
(427~mas) at Epoch~1 (Epoch~2). 

It is of interest to compare these values with the work of Diaz,
Costa \& Jatenco-Pereira (2001), who resolved the remnant of \ncas\
at 2.1\mic\ at $t=2505$~days (relative to the time origin assumed
here). They determined an angular diameter of 840~mas, and concluded
that emission by the resolved remnant they observed could not be
free-free, and that the SED of the remnant could be represented by a
blackbody at 1100~K. This value led Diaz et al. to conclude emission by
dust; however our results, as well as those of Mason et al. (1998), show
that the dust temperature was as low as $\sim600$~K at $t\simeq250$~days,
and it is difficult to see how the dust temperature could be elevated to
1100~K by $t=2505$~days.

{\sl ISO} observations of \ncas\ (Salama et al. 1999), at $t=950$,
1265, and 1455~days, failed to detect any dust, with a limit of
0.3~Jy ($\sim1.4\times10^{-13}$~W~m$^{-2}\mic^{-1}$) at $\sim2.5$\mic,
the lower limit of the {\sl ISO}\/ SWS. The flux in the resolved shell
at 2.2\mic\ was $4.1\times10^{-17}$~W~m$^{-2}\mic^{-1}$ (Diaz et al.
2001) so on this basis we can not rule out the possibility that these
authors detected the dust shell.

However if we and Diaz et al. have observed the same material, we
would expect the angular diameter at Epoch~1 (Epoch~2) to be 85~mas
(107~mas) for uniform expansion, significantly smaller than the
$\sim420$~mas implied by our {\sc dusty} fitting. 
There are two likely reasons for this discrepancy: First,
if the emission seen by Diaz et
al. is in the form of emission lines (e.g. Pa-$\beta$ $\lambda1.28$\mic,
Br-$\gamma$ $\lambda2.16$\mic), the $J$ and $K$ band fluxes would be
elevated relative to $H$, as observed by Diaz et al. For example, in
data obtained for our programme on 1996 August 22 ($t=982$~days), the
P$\beta$ and Br$\gamma$ fluxes are $1.7\times10^{-16}$~W\,m$^{-2}$ and
$3.2\times10^{-17}$~W\,m$^{-2}$ respectively. Using the filter
bandwidths for the system used by Diaz et al., together with their $J$
and $K$-band fluxes for the extended shell, their in-band fluxes are
$\sim2.9\times10^{-17}$~W\,m$^{-2}$ ($J$) and
$\sim1.4\times10^{-17}$~W\,m$^{-2}$ ($K$) respectively. Given the
difference in time between our 1996 August observation and that of
Diaz et al., and the likely decline in the line fluxes between
the observations, it seems plausible that they detected
extended emission from recombination lines.

Second, with the $T_{\rm d}$ values from Table~\ref{dustyfit}
and bolometric luminosity $L_* = 5.6\times10^4$\Lsun, blackbody
grains would lie at distance $\sim7.5\times10^{14}$~cm from the
stellar remnant, an order of magnitude smaller than the $R_{\rm in}$
values in Table~\ref{dustyfit}. On the other hand, spherical graphitic
grains of radius $a_{\rm max}$ and having the Planck mean absorption
efficiency given by Gilman (1974) would be at $\sim5\times10^{15}$~cm.
The deduced $R_{\rm in}$ depends on both the nature of the grain
material and (via the $T_*$-dependence of its Planck mean absorption
efficiency) on $T_*$; the situation is further complicated in the
case of \ncas\ by the grain mix and by the relatively weak constraint
on $T_*$ (see \S~\ref{remtemp}). This underlines the fact (see above)
that the dust shell dimensions are not uniquely determined by the
radiative transfer problem.

\subsection{The grain size distribution}
\label{size}

The grain size distribution $n(a)\,da \propto a^{-q}\,da$, with
$q\simeq2.3$, with no significant change in $q$ between the two Epochs.
In the environments of evolved stars, it is usual for the value of
$q$ to be $\gtsimeq3$, a value typical of grain-grain shattering
(Hellyer 1970; Jones et al. 1995). The size distribution we have
deduced for \ncas\ is significantly flatter than this. Such a
situation may arise if grain growth/coagulation, destruction and
shattering in the wind of \ncas\ competed in such a way that
the larger grains survived at the expense of smaller, possibly
as a result of the former sweeping up the latter. Given that
the charging of grains in nova winds has a significant impact
on the rate at which grains grow, coagulate and ablate (Shore
\& Gehrz 2004; Rawlings \& Evans 2005), this is not unexpected;
however we do not pursue this in detail here.

Evans \& Rawlings (1994) have concluded that free-flying polycyclic
aromatic hydrocarbon (PAH) molecules would not survive the harsh
nova radiation field. The sizes of the smallest grains in the
distribution are not well-constrained, so we can not confidently
determine whether the UIR features arise in free-flying PAHs or
from the vibration of C\chemone H bonds on the surface of
hydrogenated amorphous carbon (HAC) grains.

The maximum grain size, on the other hand, is rather
better constrained. We find that the largest grain size
is $\simeq0.06$\mic\ at both Epochs. This is significantly
smaller than deduced in Paper~II ($\ltsimeq0.54$\mic\ at
Epoch~1, $\gtsimeq0.57$\mic\ at Epoch~2, and asymptotic
dimensions $\sim0.7$\mic) on the basis of
$\lambda^{-\beta}B(T,\lambda)$ fits to the data; such
discrepancies are not unexpected given the simple-minded
treatment in Paper~II. However it is also significantly
smaller than the $a\gtsimeq0.2$\mic\ (at $t=65.5$~days)
deduced -- within days of dust precipitation -- by Shore
et al. (1994), on the basis of flat UV extinction
shortward of 0.28\mic, and is substantially smaller than
the grain size generally deduced in nova dust shells,
$\sim0.5-1$\mic\ (e.g. Gehrz et al. 1998). However we
have performed several tests with large initial values
of $a_{\rm max}$ and $a_{\rm max}\simeq0.06$\mic\ results
irrespective of the starting value in the fitting routine,
and we consider that this conclusion is robust.

We believe that the discrepancy between our $a_{\rm max}$ and
that deduced by Shore et al. (1994) may have arisen, in part,
for the following reason. Shore et al.'s conclusion was based
on the fact that the UV extinction was neutral shortward of
$\lambda\simeq2800$\AA; they assumed that this implies $2\pi
a/\lambda>4$, so that $a>0.18$\mic. While this is valid for
primarily scattering grains in the interstellar medium (see
Spitzer 1978), it is less so for strongly absorbing (e.g.
carbonaceous) grains in circumstellar environments, for which
the condition for neutral extinction is closer to $2\pi
a/\lambda>1$. Indeed, the IR spectroscopic evidence is that
the AC grains were the first to condense in the wind of \ncas\
(see \S\ref{comp}), so the UV extinction seen by Shore et al.
would have been due to AC rather than silicates. For \ncas\
therefore, the condition $2\pi a/\lambda>1$ is likely the more
appropriate, so that neutral extinction implies
$a\gtsimeq0.045$\mic. This would be more in line with the
$a_{\rm max}$ values deduced here and would imply that, after
an initial ($\sim10$~day) phase of very rapid increase in
grain size, grain growth levelled off, possibly as the
condensate was depleted.

More generally, grain size in nova winds is often deduced from
\begin{eqnarray}
\label{armax}
a & = & \frac{L_*}{16\pi\sigma T_{\rm d}^4 R^2 (Q_{\rm e}/a)} \:\:\: \phi(\tau_{\rm ext}) 
           \\
 & \simeq &  1.87\times10^{22} \, \left [ \frac{L_*}{\Lsun} \right ]
      \left [ \frac{V}{\vunit} \right ]^{-2} \: 
      \left [ \frac{t}{\mbox{days}} \right ]^{-2}  T_{\rm d}^{-6}  \mic \nonumber 
\end{eqnarray}
(see Gehrz et al. 1980);  this follows from equating the total
power absorbed by the grains in the shell to the total re-radiated
power. Here $L_*$ is the (constant) bolometric luminosity of the
stellar remnant, $R$ is the grain-nova distance, $Q_{\rm e}$ is the
Planck mean of the grain emissivity, $\tau_{\rm ext}$ is the optical
depth due to extinction in the dust shell and $\phi(\tau_{\rm ext})$
is a factor that takes into account extinction within the shell (see
below). Equation~(\ref{armax}) assumes graphitic grains, for which
$Q_{\rm e} \propto aT_{\rm d}^2$

Equation~(\ref{armax}) with $\phi(\tau_{\rm ext})\equiv1$ follows if
(i)~there is no internal extinction in the dust shell and (ii)~the
dust shell is geometrically thin so that the isothermal grain
approximation applies. Application of Equation~(\ref{armax}) to
novae at IR maximum leads typically to grain sizes $\sim1$\mic,
considerably larger than even the $a_{\rm max}$ deduced here. For
example, in the case of \ncas, and taking relevant parameters from
Mason et al. (1998), we find $a(\mbox{IR max}) \simeq 1.4$\mic.

A possible explanation of course is that, as the $a_{\rm max}$
deduced here is for $t>253$~days, well after IR maximum (which
occurred around $t\sim100$~days in \ncas; Mason et al. 1998),
there was a period of rapid grain destruction following
the grain growth that preceded IR maximum.
Indeed, application of Equation~(\ref{armax}) to \ncas\
for times when $\tau_{\rm ext}$ was small suggests that the
grain radius initially increased, and eventually declined.
Various scenarios for grain destruction have been proposed by
Mitchell and coworkers (Mitchell, Evans \& Bode 1983; Mitchell
\& Evans 1984; Mitchell, Evans \&  Albinson 1986) and it is
inevitable that grains in nova winds are heavily processed
(see Rawlings \& Evans 2005).

Alternatively Equation~(\ref{armax}) may not be appropriate at
IR maximum for a dust shell that is {\em optically thick at short
wavelengths:} as we have seen, assumption (ii)~above is a good
approximation in the case of \ncas\ but (i)~is not. In this case,
\[ \phi(\tau_{\rm ext}) = \frac{1-\exp[-\tau_{\rm ext}]}{\tau_{\rm ext}} \:\: , \]
which $\rightarrow1$ as $\tau_{\rm ext} \rightarrow 0$. Applying
this correction to \ncas\ at IR maximum gives $a(\mbox{IR max})
\simeq 0.2$\mic\ for $\tau_{\rm ext} \simeq 5.5$ (see
\S\ref{thickness}), considerably less than that obtained on the
assumption that the dust shell is optically thin.

We also note that Equation~(\ref{armax}), with
$\phi(\tau_{\rm ext})\equiv1$, will apply whenever $L_*$ is
constant and the dust shell is optically thin at short wavelengths (as
is the case after light curve recovery), as in \ncas\
at Epochs~1 and~2. Using
$L_*=5.6\times10^4$\Lsun\ from \S\ref{depth} and $V=850$\vunit,
Equation~(\ref{armax}) with $\phi(\tau_{\rm ext})=1$ gives 0.5\mic\
(0.3\mic) at $t=253$~days (320~days), again considerably smaller
that that obtained for the time of IR maximum.
While the situation in \ncas\ is complicated by the presence of (at
least) two grain types, we conclude that the application of
Equation~(\ref{armax}) to determine grain size is inappropriate at
infrared maximum in novae having optically thick dust shells.

\subsection{The dust composition}
\label{comp}

We find that both sets of data are best fitted by a dust shell in
which carbon grains are numerically more abundant than silicate
grains. We do not expect the AC:silicate fraction to vary as the
dust shell moves away from the stellar remnant, and the individual
values of the AC and silicate fractions in Table~\ref{dustyfit}
are gratifyingly constant; for the purpose of discussion we average
the AC:silicate ratio, and integrate over the grain size distribution
(see \S\ref{size} below) to get a $\sim0.67:0.33$ ($\pm\sim0.1$)
carbon:silicate mix. This translates to a mass ratio carbon:silicate
= 1.26, significantly lower that the carbon:silicate ratio of
$\sim8.5$ obtained by Mason et al. (1998). However these authors
fit a 4-component model to limited broadband data, with UIR
components at their `standard' wavelengths, and they caution that
their derivation of the silicate mass may be problematic. We
should not therefore be surprised by the discrepancy and consider
that the result obtained here is reasonably robust.

The co-existence of carbon and silicate dust (`chemical dichotomy')
in the IR spectra of evolved stars is usually ascribed to the trapping
of silicate dust in a disc arising from mass-loss during an earlier
phase of evolution when the star was oxygen-rich, while carbon dust
arises from more recent or current (carbon-rich) mass-loss (see e.g.
Zijlstra et al. 2001 and references therein). In the case of \ncas,
UIR emission was present in the 10\mic\ band when silicate was weak
or absent (Evans et al. 1997). Novae may also have two well-separated
mass-loss episodes with different chemistries to produce their chemical
dichotomies, or there may be steep abundance gradients in the ejecta,
or (as CO formation does not go to saturation; see Paper~I, 
Pontefract \& Rawlings 2004, Rawlings \& Evans 2005) the nucleation
chemistry allows the simultaneous formation of both carbon and silicate 
grains. In any of these cases, there may be interesting implications for 
the thermonuclear runaway, the nature of the nova explosion and the 
pre-dust chemistry.

While the assumptions leading to Equation~(\ref{tcond}) may not
necessarily apply in the case of novae (see Evans \& Rawlings
1994, Rawlings \& Evans 2005 for a discussion of this), our values
for the dust temperature at the inner edge of the shell imply
condensation temperature $T_{\rm c}\simeq1060^{+5}_{-28}$~K
($1153^{+23}_{-46}$~K) for Epoch~1 (Epoch~2); again the value
of $T_{\rm c}$ is not sensitive to $\beta$ and we have assumed
$\beta=1$. These values are of the right order for carbon
condensation; however the situation in \ncas\ is complicated
by the presence of two grain types, so the meaning of the
`condensation temperature' is not immediately obvious in this
case. On the other hand, UIR emission and an underlying
continuum, but no 9.7\mic\ silicate feature, were present in
1994 May (Paper~II), implying that the AC condensed before the
silicate.

\subsection{The dust features}

\subsubsection{The silicate features}

\begin{figure}
\begin{center}
\setlength{\unitlength}{1cm}
\begin{center}
\leavevmode
\begin{picture}(6.0,5.5)
\put(0.0,4.0){\includegraphics{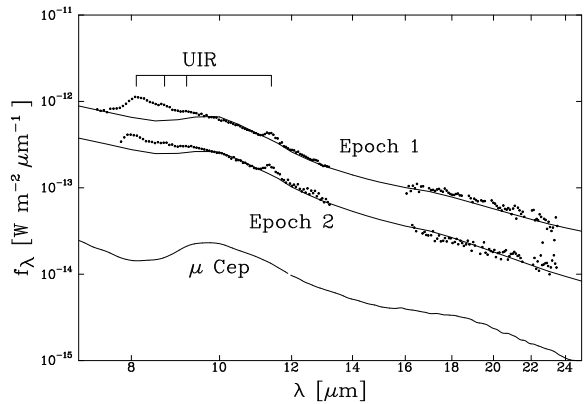}}
\end{picture}
\end{center}
\caption[]{The profiles of the silicate features in \ncas\ for the two
Epochs; the wavelengths of the 8.1, 8.6, 9.2 and 11.4\mic\ UIR features
(see Fig.~\ref{UIR}b and Table~\ref{uirs}) are indicated. The
curves through the data are the {\sc dusty} fits; see text for details.
Also shown for comparison (lower curve) is the {\sl ISO} SWS spectrum
of the red supergiant $\mu$~Cep.}
\label{cas_sil}
\end{center}
\end{figure}

In Paper~II we suggested that, in addition to the 9.7\mic\ silicate
feature, there might also be present the corresponding 18\mic\ feature.
The IR spectrum, together with the fitted model, in the range 7-25\mic\
for the two epochs is shown in Fig.~\ref{cas_sil}; it seems that the
18\mic\ feature may indeed be present in the IR spectrum of \ncas, as
may be seen by viewing Fig.~\ref{cas_sil} at grazing incidence.

As noted in Paper~II, Nuth \& Hecht (1990) suggested, on the basis
of laboratory experiments on silicate analogues, that the strength
of the 18\mic\ silicate feature relative to that of the 9.7\mic\
feature increases as the silicate `ages' and is subjected to
annealing. Processing of silicates, and its effect on the
9.7\mic\ feature in circumstellar environments, has been
discussed from an observational point of view by Bouwman et
al. (2001); however the objects (Herbig Ae/Be stars) in this
study are young objects, and are not themselves dust producers
and comparison with the silicates around evolved, dust-forming
stars, is more relevant from our point of view.

To explore this we also include in Fig.~\ref{cas_sil} the {\sl ISO}
SWS spectrum of the actively dust-forming red supergiant $\mu$~Cep,
which also displays prominent silicate features (see Tielens et al.
1997). We note in particular
\begin{enumerate}
\itemsep=2mm
\item the similarity of the 9.7\mic\ feature profiles in the two
objects. The features are narrow, structureless, and peak around
9.7\mic, suggestive of amorphous silicate;
\item the relative strength of the 9.7\mic\ and 18\mic\ features,
in that the ratio 18/9.7 is much smaller in \ncas\ than in $\mu$~Cep,
consistent with the view that the silicate in the nova is much
`fresher' than that in $\mu$~Cep.
\end{enumerate}
While our data lend support to the conclusion of Nuth \& Hecht
(1990), the mid-IR spectra of novae, obtained over periods ranging
from $\sim20$ to $\sim1000$~days from outburst (and hence over a
range of annealing times), suggests that the situation is more
complex than this. Although \ncas\ is now a very weak mid-IR
source (the dust was not detected with the {\sl ISO} SWS by
Salama et al. 1999), further observations in the mid-IR (e.g.
with the {\sl Spitzer} Space Telescope (Werner et al. 2004))
will be valuable to monitor both the profile of the 9.7\mic\
feature -- which may shift to longer wavelengths and display
structure as the dust anneals -- and the development of the
relative strengths of the 9.7\mic\ and 18\mic\ features.

\begin{figure*}
\begin{center}
\setlength{\unitlength}{1cm}
\begin{center}
\leavevmode
\begin{picture}(6.0,5.5)
\put(0.0,4.0){\includegraphics{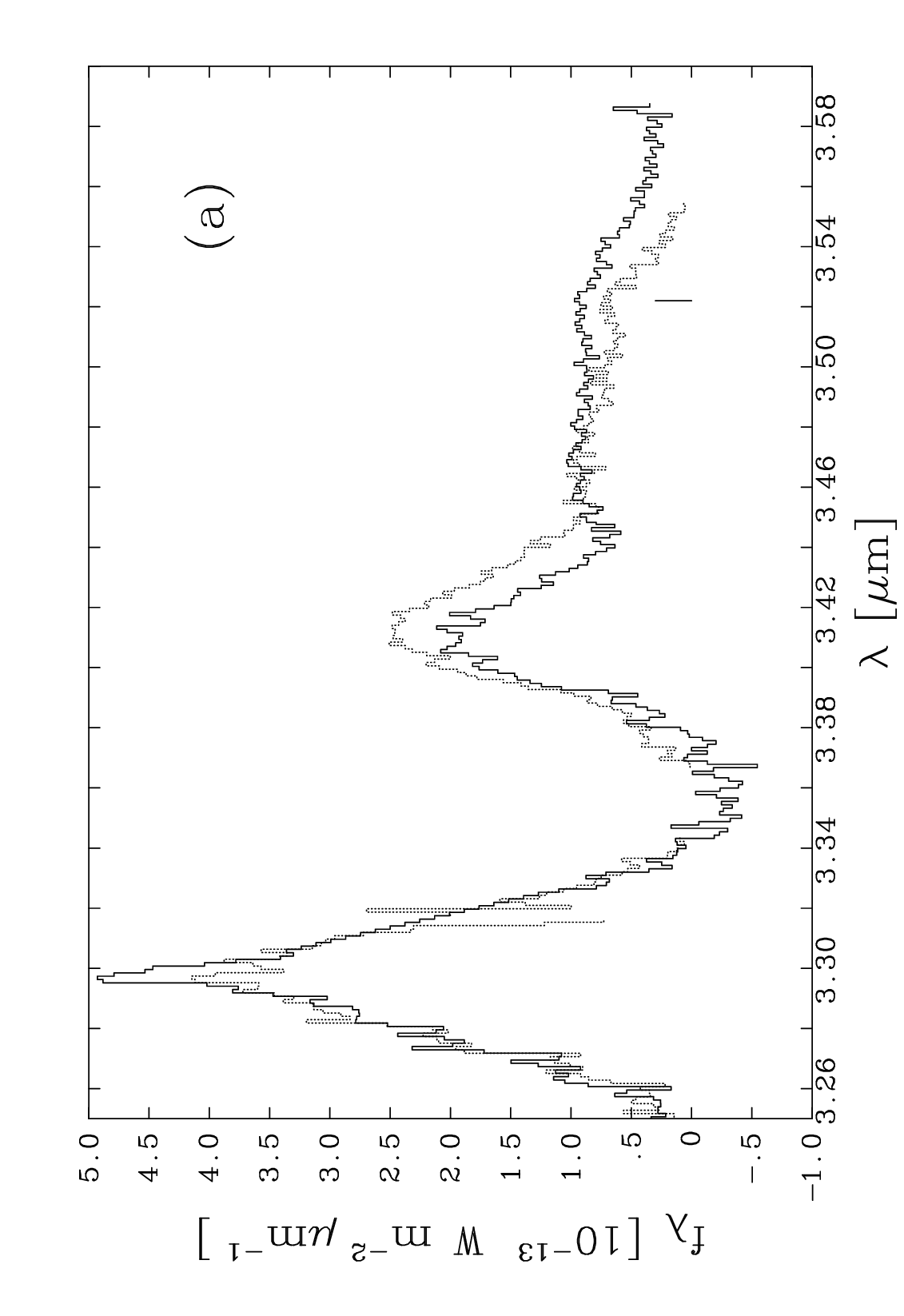}}
\put(0.0,4.0){\includegraphics{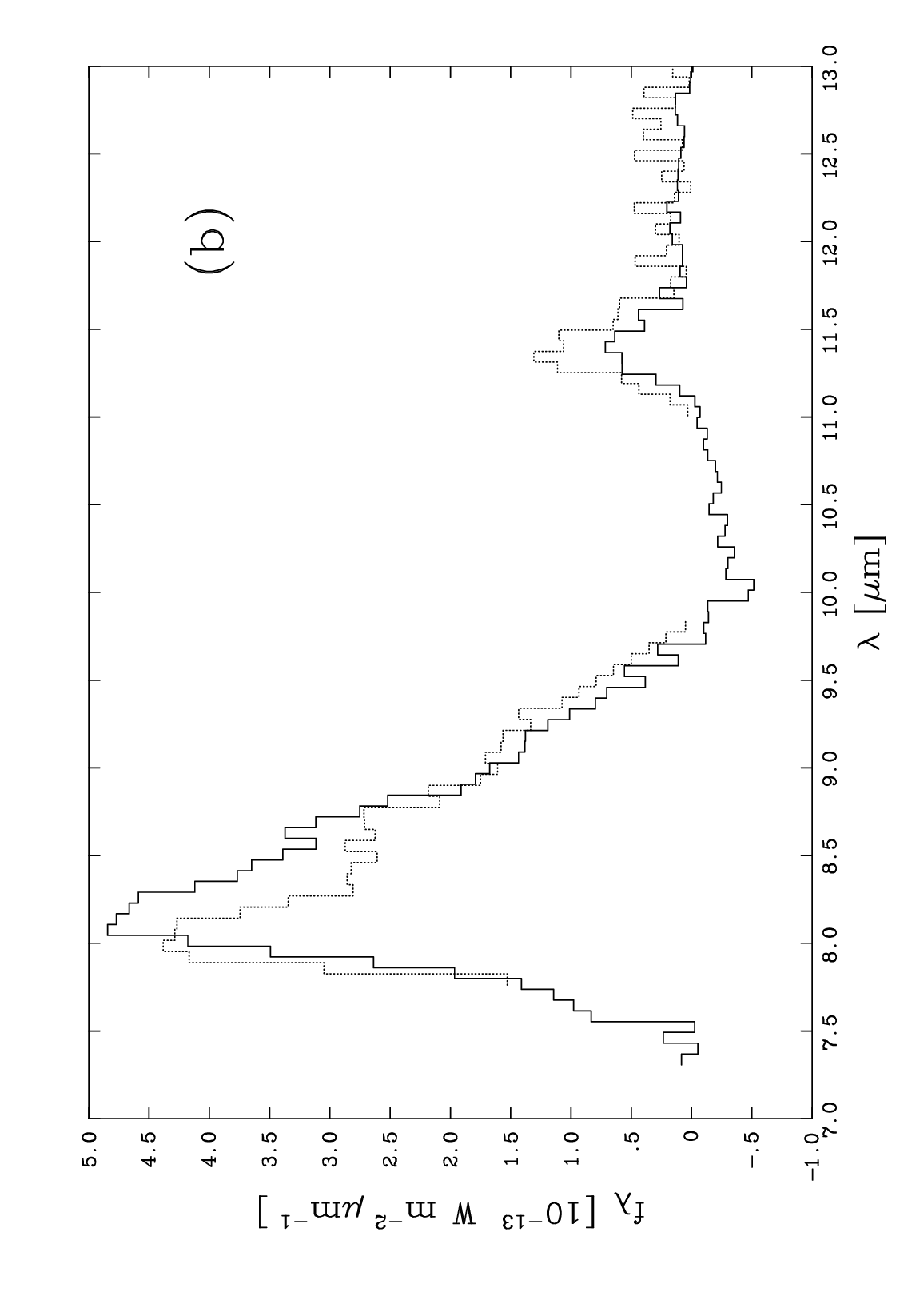}}
\end{picture}
\end{center}
\caption[]{The profiles of the UIR features in \ncas; solid line, Epoch 1,
broken line, Epoch 2. (a) The $L$-band features; the tick mark indicates
the feature at 3.522\mic\ (see text). (b) the $N$-band features. Note the
shift in wavelength of the 8\mic\ feature. See text for details.}
\label{UIR}
\end{center}
\end{figure*}

\begin{table*}
\begin{center}
\caption{Properties of the UIR features in the $N$-band.}
\label{uirs}
\begin{tabular}{cccclc}
Epoch & $\lambda$  & FWHM        & Flux   & \multicolumn{1}{c}{Identification}
                   & Standard $\lambda$ ($\!\mic$)\\ 
      & ($\!\mic$) &  ($\!\mic$) &  ($10^{-14}$~W\,m$^{-2}$) & & for Class A UIRs \\ \hline
1  & $8.17\pm0.01$  & $0.63\pm0.03$ & 32.1 ($\pm1.4$)  & Si\chemone CH$_3$ vibration?  & \ldots \\ 
   & $8.72\pm0.02$  & $0.41\pm0.03$ & 10.8 ($\pm8.1$)  & NH$_2$ rock? & \ldots \\  
   & $9.20\pm0.02$  & $0.46\pm0.04$ & 6.27 ($\pm0.43$) & ? & \ldots \\  
   & $11.41\pm0.04$ & $0.34\pm0.08$ & 2.53 ($\pm0.51$) & Aromatic C\chemone H bend  & 11.3\\ 
   &      &            &                 &  & \\ 
2  & $8.06\pm0.06$  & $0.47\pm0.21$ & 7.27 ($\pm2.46$) & Si\chemone CH$_3$ vibration? & \ldots \\  
   & $8.67\pm0.08$  & $0.53\pm0.20$ & 4.86 ($\pm1.89$) & NH$_2$ rock? & \ldots \\  
   & $9.26\pm0.15$  & $0.52\pm0.24$ & 2.34 ($\pm1.29$) & ? & \ldots \\  
   & $11.39\pm0.12$ & $0.38\pm0.19$ & 1.51 ($\pm1.14$) &  Aromatic C\chemone H bend  & 11.3 \\ \hline
\end{tabular}
\end{center}
\end{table*}

\subsubsection{The UIR features}
\label{UIRf}
We have subtracted the model fit from the data in the $L$ and $N$
bands to reveal the profiles of the UIR emission (see Fig.~\ref{UIR}a,b).
Clearly the resultant UIR profiles in the $N$ band depend to some
extent on the profile of the warm silicate feature (Ossenkopf et
al. 1992) but, as we are looking for possible changes in the UIR
features this does not qualitatively affect the discussion that
follows.

In Fig.~\ref{UIR}a the $L$-band data from Epoch~2 data have been
normalized to the Epoch~1 data at 3.4\mic\ and it is evident that,
while the relative strengths of the 3.28\mic\ feature and the
3.5\mic\ `plateau' are essentially unchanged between the two
Epochs, the 3.4/3.28 ratio may have changed. 

There is a weak feature at $\lambda=3.522\pm0.001$\mic\ (see
Fig~\ref{UIR}a). While this is close to the UIR feature at
3.52\mic\ (see Geballe 1997), the width of the feature in \ncas\
($\mbox{FWHM} \sim 0.014$\mic\ at Epoch~2 and corresponding to
an expansion velocity $\sim600$\vunit) suggests that it might
be a nebular feature rather than a dust feature, which would
typically be much broader than this (see Table~\ref{uirs}
below). Further IR spectroscopy of dusty novae in this spectral
region is desirable to resolve this issue.

A possible (nebular) identification for this feature is
\pion{H}{i} 23-6 $\lambda3.522$\mic, although there is no evidence
for \pion{H}{i} 24-6 $\lambda3.501$\mic, which is expected to be
of comparable strength. An alternative identification, particularly
in view of the common occurrence of \fion{Fe}{ii} lines in the IR
spectra of novae (Rudy et al. 2003; Evans et al. 2003), is
\fion{Fe}{ii} $a^6$D$_{9/2} - a^4$F$_{5/2}$ at $\lambda=3.52367$\mic.
However there are many permitted and forbidden \pion{Fe}{ii}
transitions in the wavelength range for which we have data and we
defer a discussion of the emission line spectrum of \ncas\ to a
later paper. 

In Fig.~\ref{UIR}b the $N$-band data from Epoch~2 data have been
normalized to the Epoch~1 data at 9\mic; the primary feature at
$\sim8.1$\mic\ has a `shoulder' at $\sim8.6$\mic\ and $\sim9.2$\mic.
We have fitted the `8.1\mic' feature with 3 gaussians (see
Fig.~\ref{UIR10}a,b and Table~\ref{uirs}). We note from Table~\ref{uirs}
that the widths of the features are typically $\sim0.3-0.6$\mic,
and so these are unlikely to be nebular in origin. Also listed in
Table~\ref{uirs} are the `standard' wavelengths and likely identification
of `Class~A' UIR features (i.e. those usually seen in planetary nebulae
and \pion{H}{ii} regions) from Geballe (1997).

There is no evidence that the central wavelengths of the two
`shoulder' features ($\sim8.72$\mic, $\sim9.20$\mic) change between
Epochs~1 and 2; however there seems to be a distinct change in the
peak wavelength of the primary feature between Epochs~1 and~2 (see
Fig.~\ref{UIR}a and Table~\ref{uirs}), from 8.17\mic\ to 8.06\mic.
This is not an artefact of the subtraction of the model from the
data, and likely points to a change in the character of the carrier.

As noted in Paper~II, the 11.3\mic\ feature in \ncas\ appears at a
longer wavelength than is usual (the slight difference between the
values in Table~\ref{uirs} and in Paper~II is a result of the slight
difference in the subtracted continuum). We note that there seems to
be a significant increase in the strength of the 11.3\mic\ feature
relative to that of the `8.1' feature.

\subsubsection{Why are nova UIR features unique?}

It is of interest to consider why nova UIR features are in a class of
their own (Geballe 1997), and in particular are significantly different
from those seen in other astrophysical environments.

Laboratory measurements have shown that the peak wavelengths of the
`3.28', `6.2', `7.7', `8.1' and `11.3' UIR features in quenched
carbonaceous composite depend on the $^{13}$C/$^{12}$C ratio, the
peak wavelengths increasing linearly with the $^{13}$C/$^{12}$C
ratio (Wada et al. 2003). However there are two reasons why this
mechanism can not be operating in \ncas. First, the $^{13}$C/$^{12}$C
ratio in \ncas\ was low, $\ltsimeq0.2$ (Paper~I), so isotopic effects
will have negligible effect on the wavelengths of the UIR features;
however this effect must clearly be looked for in novae with high
$^{13}$C/$^{12}$C ratio. Second, the peak wavelengths of the nova
3.28 and 3.4 features are not anomalous, only their relative
strengths are unusual. If isotopic effects were important the peak
wavelengths of all the UIR features would be affected, which was
not the case in \ncas.

\begin{figure*}
\begin{center}
\setlength{\unitlength}{1cm}
\begin{center}
\leavevmode
\begin{picture}(6.0,5.5)
\put(0.0,4.0){\includegraphics{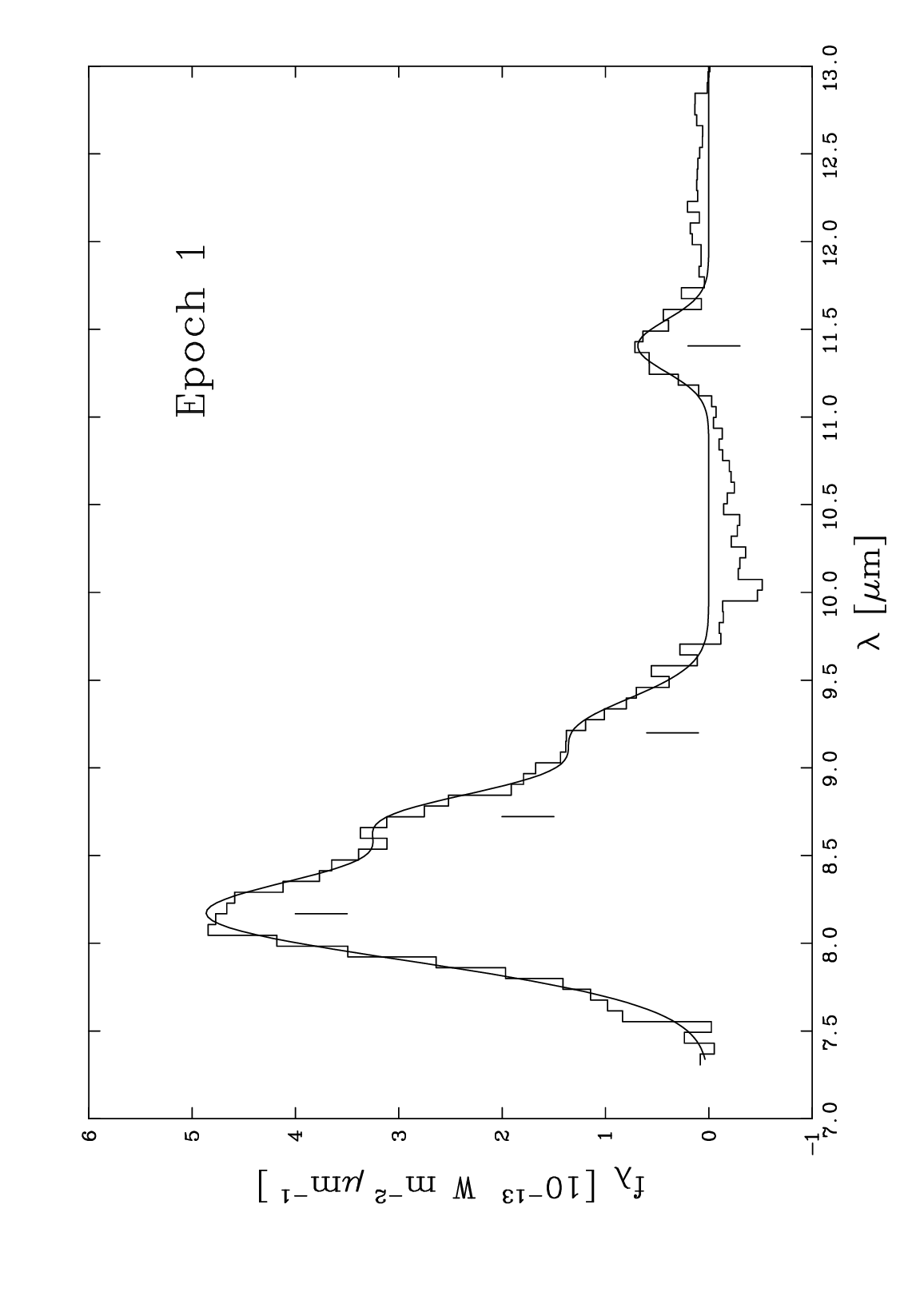}}
\put(0.0,4.0){\includegraphics{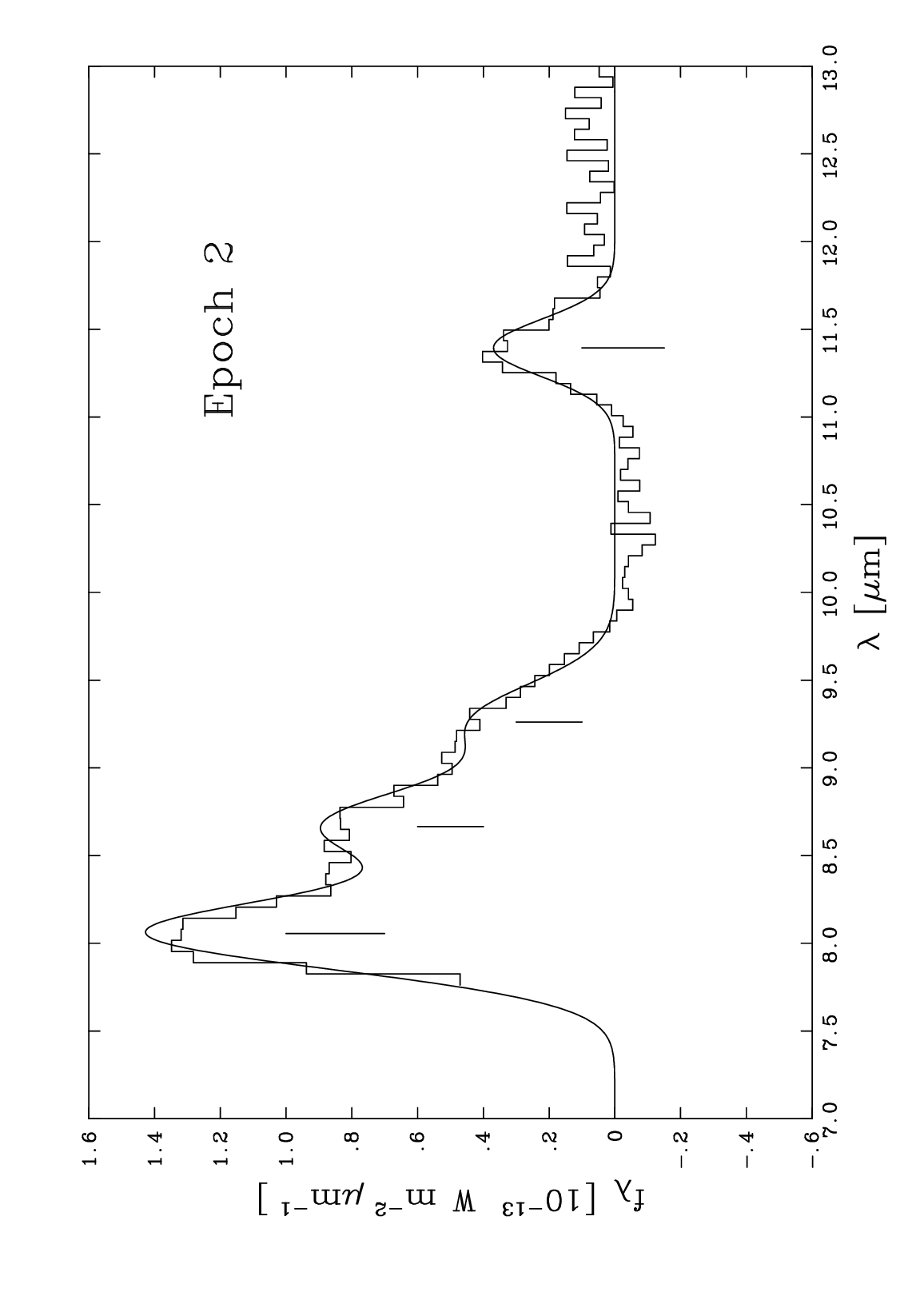}}	 
\end{picture}
\end{center}
\caption[]{The profiles of the 7-13\mic\ UIR features in \ncas\ for Epoch~1
(left) and Epoch 2 (right). Curves are fitted gaussians; ticks marks locate
peaks of individual features. See also Table~\ref{uirs}.}
\label{UIR10}
\end{center}
\end{figure*}

More plausibly, the nature of the UIR carrier in novae is governed
by the unique environment in which the pre-dust and dust chemistry
takes place. We consider separately the UIR features in the 3\mic\
and 7-13\mic\ windows. 

\paragraph{The 3\mic\ window.}

As is well known, the relative strengths of the `3.28' and `3.4'
features in novae differ substantially from those in other sources
such as planetary nebulae and post-AGB stars (see e.g. Geballe 1997,
van Diedenhoven et al. 2004). A possible explanation for this is the
incorporation of CH$_2$ and CH$_3$ groups in the silicate lattice,
which experimental studies (Grishko \& Duley 2002a) have shown
enhances the 3.4\mic\ UIR feature.

As noted in \S\ref{comp}, one explanation for the chemical dichotomy
in nova dust shells is that CO does not proceed to saturation,
allowing the simultaneous condensation of carbonaceous and silicate
dusts: the chemistry of grain formation and growth is likely to
include the formation of carbonaceous grains with silicate inclusions,
and vice versa. The incorporation of CH$_2$/CH$_3$ groups into the
silicate lattice in this way is consistent with the simultaneous
presence of C-bearing and Si/O-bearing dust, and underlines the
irrelevance of the C:O ratio to the nature of the condensate in
nova winds.

Furthermore, (see \S\ref{size}), the grain size distribution
indicates that larger grains may have swept up the smaller. Unless
there were (for example) substantial species-dependent grain charge
effects it is unlikely that larger grains would discriminate between
one grain type and another in the sweeping-up process; this would
also give rise to the kind of composite grain hinted at by the
3\mic\ UIR features.

\paragraph{The 7-13\mic\ window.}

Additional support for the hypothesis that CH$_2$/CH$_3$ inclusions
in silicates could be significant in novae comes from the `8.1'
feature in \ncas. A feature at this wavelength is also seen in
laboratory spectra, and may be identified with Si\chemone CH$_3$
vibration (W. W. Duley, private communication). Indeed Duley
(private communication) has suggested that the shift in the peak
wavelength of the `8.1' feature in \ncas\ between Epochs~1 and 2
(see Fig.~\ref{UIR} and Table~\ref{uirs}) may be the result of a
change in the co-ordination of Si\chemone CH$_3$ groups.

Furthermore, it is likely that the UIR carrier in \ncas\ (and
indeed in other novae) is heavily contaminated by other species,
notably O or N, particularly the latter given its extreme
overabundance ($\sim100$; see summary in Gehrz et al., 1998)
relative to solar in nova winds.
Grishko \& Duley (2002b) have investigated experimentally
the IR spectra of HAC prepared in the presence of various
contaminants, and find that HAC deposited in the presence
of N-bearing molecules (particularly NH$_3$) displays several
features not present in `normal' UIR carriers. In particular,
there is a feature at 8.56\mic\ (which Grishko \& Duley attribute
to NH$_2$ rock), and a broad (unidentified) feature centred at
$\sim9.1$\mic\ with width $\sim0.4$\mic, which are close to the
features on the `shoulder' of the 8.1\mic\ feature in \ncas\
(see Fig.~\ref{UIR10}a,b and Table~\ref{uirs}). We tentatively
suggest that the carrier of the UIR features in \ncas\ had a
significant nitrogen component.

\section{Conclusions}

We have used the {\sc dusty} code to reassess the properties of the
optically thick dust shell around the classical nova V705~Cas.
We find
\begin{enumerate}
\item the dust shell is a $\sim2:1$ (by number) mix of amorphous
carbon and `warm silicate';
\item the grain size distribution $n(a)\,da \propto a^{-q}\,da$
was flat, with $q\simeq2.3$; this may indicate that the larger grains
swept up the smaller grains;
\item the maximum grain size in the distribution was $a_{\rm max}
\simeq 0.06$\mic, considerably less than that deduced from previous
studies (Shore et al. 1994, Paper~II, Gehrz et al. 1998);
\item the dust-bearing shell was, like the CO-bearing shell,
geometrically thin, such that $\Delta R/R \simeq 0.1$;
\item the condensation temperature of the dust was $\sim1100$~K,
although in \ncas\ the situation is complicated by the presence
of two condensates;
\item structure in the 8-9\mic\ UIR complex may be due to
nitrogenation of the UIR carrier;
\item the strength of the 3.4\mic\ feature relative to the 3.28\mic\
feature, commonly seen in novae, as well as the 8.1\mic\ feature,
may be due to the incorporation of CH$_2$ and CH$_3$ groups in the
silicate matrix;
\item the silicate dust at least was freshly condensed, judging by
the weakness of the 18\mic\ silicate feature;
\item the extrapolated angular diameter of the dust shell is such that
the resolved remnant seen by Diaz et al. (2001) was very likely due
to line emission rather than dust.
\end{enumerate}

\section*{Acknowledgments}
We thank Profs~W. Duley and R. Gehrz
for commenting on an early draft of this paper.
TRG is supported by the Gemini Observatory, which is operated by the
Association of Universities for Research in Astronomy, Inc., on behalf
of the international Gemini partnership of Argentina, Australia,
Brazil, Canada, Chile, the United Kingdom, and the United States of
America. SPSE acknowledges support from the Nuffield Foundation.
VHT was supported by Keele University.

\bsp

\label{lastpage}

\end{document}